\documentclass[reprint,superscriptaddress,amsmath,amssymb,aps,prb,showpacs]{revtex4-1}
\usepackage{graphicx,dcolumn,bm,natbib,color}
\newcommand\HoB{HoB$_4$}
\newcommand\RB{$R$B$_4$}
\newcommand\MMF{$M/M_{\mathrm{50kOe}}$}
\newcommand\muB{$\mu_{\mathrm{B}}$}

\newcommand\afm{antiferromagnetic}
\newcommand\TNL{$T_{\mathrm{N2}}=5.7$~K}
\newcommand\TNH{$T_{\mathrm{N1}}=7.1$~K}

\begin{document}
\preprint{APS/123-QED}
\title{Field Induced Magnetic States in Holmium Tetraboride}
\author{D. Brunt}
\author{G. Balakrishnan}
\affiliation{Department of Physics, University of Warwick, Coventry CV4 7AL, United Kingdom}
\author{A.R. Wildes}
\author{B. Ouladdiaf}
\author{N. Qureshi}
\affiliation{Institut Laue-Langevin, 6 rue Jules Horowitz, BP 156, 28042 Grenoble Cedex 9, France}
\author{O. A. Petrenko}
\affiliation{Department of Physics, University of Warwick, Coventry CV4 7AL, United Kingdom}

\date{\today}
\begin{abstract}
A study of the zero field and field induced magnetic states of the frustrated rare earth tetraboride, \HoB, has been carried out using single crystal neutron diffraction complemented by magnetisation measurements.
In zero field, \HoB\ shows magnetic phase transitions at \TNH\ to an incommensurate state with a propagation vector $(\delta, \delta, \delta^\prime)$, where $\delta = 0.02$ and  $\delta^\prime = 0.43$ and at \TNL\ to a non-collinear commensurate \afm\ structure.
Polarised neutron diffraction measurements in zero field have revealed that the incommensurate reflections, albeit much reduced in intensity, persist down to 1.5~K despite \afm\ ordering at 5.7~K.
At lower temperatures, application of a magnetic field along the $c$~axis initially re-establishes the incommensurate phase as the dominant magnetic state in a narrow field range, just prior to \HoB\ ordering with an \emph{up-up-down} ferrimagnetic structure characterised by the  $(h\,k\,\frac{1}{3})$-type reflections between 18 and 24~kOe.
This field range is marked by the previously reported $M/M_{\mathrm{sat}}=\frac{1}{3}$ magnetisation plateau, which we also see in our magnetisation measurements.
The region between 21 and 33~kOe is characterised by the increase in the intensity of the \afm\ reflections, such as (100), the maximum of which coincides with the appearance of the narrow magnetisation plateau with $M/M_{\mathrm{sat}}\approx\frac{3}{5}$.
Further increase of the magnetic field results in the stabilisation of a polarised state above 33~kOe, while the incommensurate reflections are clearly present in all fields up to 59~kOe.
We propose the $H-T$ phase diagram of \HoB\ for the $H\parallel c$ containing both stationary and transitionary magnetic phases which overlap and show significant history dependence.
\end{abstract}

\maketitle

\section{Introduction} \label{sec: Intro}

In geometrically frustrated systems the competing magnetic interactions lead to a large ground state degeneracy.
This suppresses long range order and often results in the formation of unusual ordered phases.~\cite{1994_Ramerez}
Application of a magnetic field can lift the degeneracy and give rise to a rich variety of new magnetic phases and in some notable examples, including garnets,\cite{1994_Schiffer_GGG,1999_Petrenko_GGG} honeycomb lattices,\cite{2010_Matsuda_honeycomb,2011_Ganesh_honeycomb} and zig-zag ladder compounds,\cite{2013_Cheffins_SDO,2016_Bidaud_SDO} induce long-range order, where none was observed down to the lowest temperatures.

\begin{figure}[tb]
\begin{center}
\includegraphics[width=0.99\columnwidth]{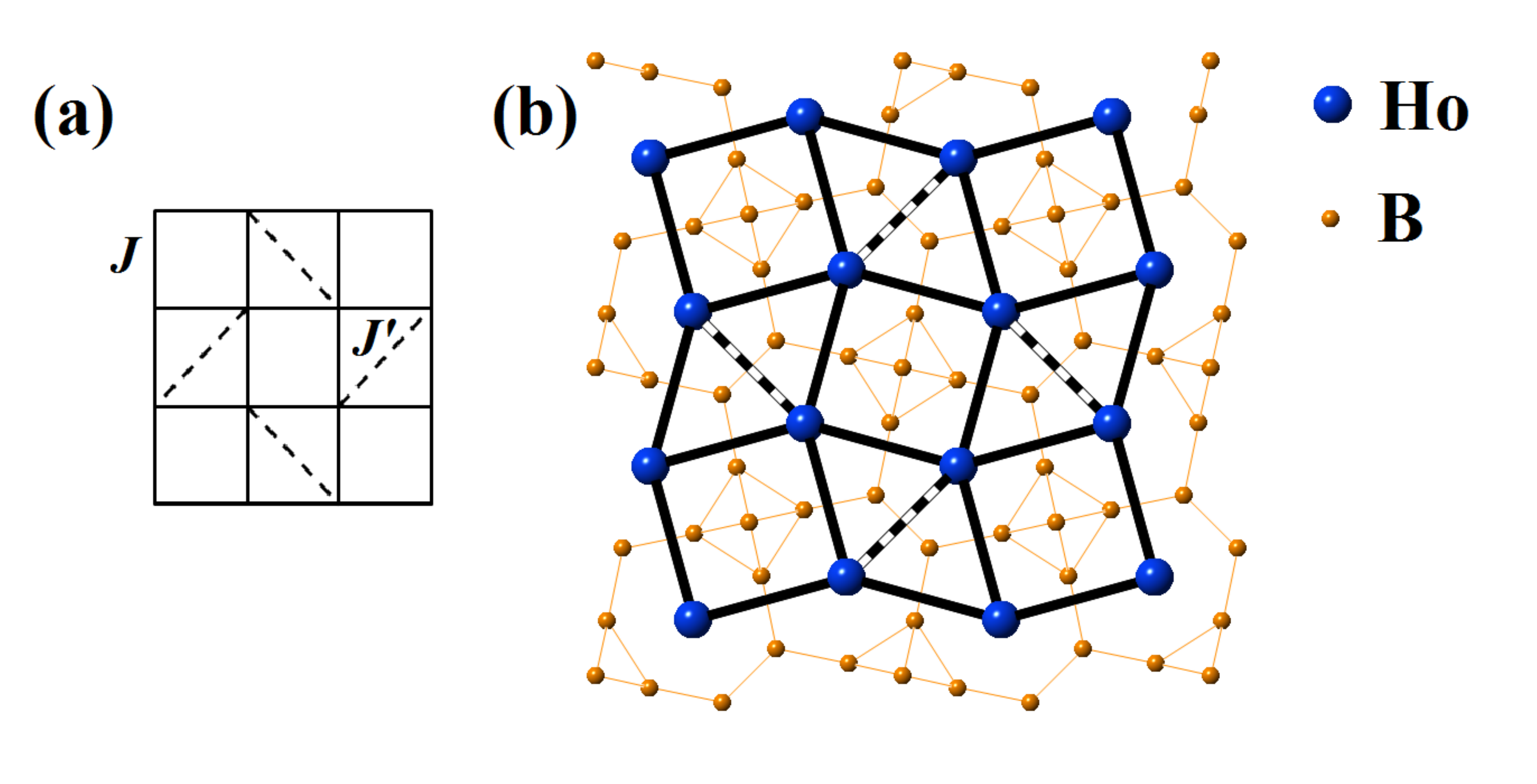}
\caption{(Colour online) (a) The Shastry-Sutherland lattice, (b) projection of \HoB\ structure along the tetragonal $c$~axis. The Ho ions form a lattice that topologically maps to the Shastry-Sutherland lattice.
Solid bonds correspond to the nearest neighbour interaction $J$, while the dashed bonds are the next nearest neighbour interaction $J^\prime$.}
\label{fig: Crystal}
\end{center}
\end{figure}

The Shastry-Sutherland lattice (SSL) is an example of a frustrated system with an exact ground state solution.
It is a square lattice with \afm\ nearest neighbour interaction, $J$ and alternating diagonal \afm\ next nearest neighbour interaction $J^\prime$ (see Fig.~\ref{fig: Crystal}(a)).\cite{1981_ShastrySutherland}
There are only a handful of experimental realisations of this lattice, the most prominent being $\rm SrCu_2(BO_3)_2$.
One of the most striking features of $\rm SrCu_2(BO_3)_2$ are the magnetisation plateaux occurring at fractional values of the saturation magnetisation, $M_{\mathrm{sat}}$.
These appear at $M/M_{\mathrm{sat}}=\frac{1}{8}$, $\frac{1}{4}$ and $\frac{1}{3}$~\cite{SCBO_Review, 1999_Kageyama_SCBO} and have been suggested to arise due to the localisation of a triplet excitation in field.\cite{1991_Miyahara_SCBO}

The rare earth tetraborides ($R$B$_4$) are another example of  the SSL.
The $R$B$_4$ family crystallises into a tetragonal structure (space group $P4/mbm$), where the $R$ ions form a network of squares and triangles in the basal plane, which maps to the SSL.
The boron atoms form chains of octahedra that run parallel to the $c$~axis.
A projection of the $ab$~plane looking along the $c$~axis is shown in Fig.~\ref{fig: Crystal}(b).\cite{1979_Etourneau}
Fractional magnetisation plateaux are also a common feature through the series.\cite{2006_Watanuki_DyB4, 2007_Iga_TmB4, 2008_Yoshii_TbB4}
These arise due to a diverse range of magnetic phases stabilised by a magnetic field across the family of compounds, including a ferrimagnetic state corresponding to $M/M_{\mathrm{sat}}=\frac{1}{2}$ in ErB$_4$,\cite{1979_Pfeiffer_ErB4} while TmB$_4$ has a complex striped structure consisting of regions of \afm\ order, separated by ferromagnetic stripes 7 or 9 units cells apart corresponding to $M/M_{\mathrm{sat}}=\frac{1}{7}$ and $\frac{1}{9}$ respectively.
An additional plateau at $M/M_{\mathrm{sat}}=\frac{1}{2}$ arises due to equally sized \afm\ and ferromagnetic stripes.\cite{2008_Siemensmeyer_TmB4}

\HoB\ shows successive magnetic phase transitions at \TNH\ and \TNL.\cite{2009_Kim_HoB4}
The magnetic structures of both the low and intermediate temperature magnetic phases have been established by powder neutron diffraction.\cite{2008_Okuyama_HoB4}
The intermediate temperature phase (abbreviated to the IT phase in the rest of the paper) was determined to be an incommensurate state with propagation vector, $\mathbf{q}=(\delta, \delta, \delta^\prime)$, where $\delta=0.022$ and $\delta^\prime=0.43$, while the low temperature phase was determined to have non-collinear \afm\ order.\cite{2008_Okuyama_HoB4}
Magnetisation measurements at $T=2$~K with a magnetic field applied parallel to the $c$~axis have shown a plateau at $M/M_{\mathrm{sat}}=\frac{1}{3}$, with further, less pronounced features occurring at $M/M_{\mathrm{sat}}\approx\frac{4}{9}$ and $\frac{3}{5}$.\cite{2010_Matas}
Significant theoretical work using Heisenberg~\cite{2009_Moliner, 2013_Huo} and Ising~\cite{2012_Huang, 2013_Dublenych, 2013_Grechnev} spins has been used to model a variety of members of the \RB\ family with many finding a plateau occurring at $M/M_{\mathrm{sat}}=\frac{1}{3}$, which is experimentally observed in TbB$_4$ and \HoB.\cite{2007_Yoshii_TbB4, 2010_Matas}
These models are generally based around the triangles formed by the diagonal $J^\prime$ coupling and include an \emph{up-up-down} structure, in which each triangle has a collinear arrangement of the spins/moments with two of them pointing along the magnetic field and one opposite to it.
Another suggestion is an ``umbrella'' structure where the moments on the triangle are rotated 120$^{\circ}$ to each other, tilted towards the vertical axis to form a spiral like structure, which propagates through the lattice.\cite{2009_Moliner, 2013_Grechnev}

A thorough understanding of the field induced behaviour in $R$B$_4$ is needed to distinguish between different theoretical models describing the magnetisation plateaux.
In this paper we report the results of our investigation into the nature of the zero-field and field induced magnetic states in \HoB\ using single crystal neutron diffraction as well as complementary magnetisation measurements.
We find that the IT incommensurate phase ``freezes'' at low temperature and persists down to at least 1.5~K.
For $H \parallel c$, this incommensurate state is also re-established as the dominant magnetic state at lower temperatures for a narrow field range, while the relatively wide $M/M_{\mathrm{sat}}=\frac{1}{3}$ plateau was found to have an \emph{up-up-down} ferrimagnetic state characterised by the $(h\,k\,\frac{1}{3})$-type reflections.
We also show that the nature of the much narrower $M/M_{\mathrm{sat}}\approx\frac{3}{5}$ plateau observed in higher fields is completely different, as it is characterised by the reappearance of an in-plane \afm\ component.  

The structure of the paper is as follows.
The experimental details are described in Sec.~\ref{Sec_Exp_Details}.
The experimental results are presented in Sec.~\ref{Sec_Results}, which is split into two subsections presenting the zero-field polarised neutron diffraction (Sec.~\ref{SubSec_D7}) and applied field, unpolarised neutron experiments with the additional magnetisation measurements (Sec.~\ref{SubSec_field}).
Finally we conclude the paper in Sec~\ref{Sec_Conclusions}.

\section{Experimental Details} \label{Sec_Exp_Details}

Polycrystalline rods of \HoB\ were prepared by arc melting the constituent elements in an argon atmosphere and the floating zone technique was then utilised to grow single crystals. Isotopically enriched boron, $^{11}$B (99\%), was used to reduced neutron absorption.
The crystals were checked and aligned using a backscattering x-ray Laue system.

Neutron diffraction measurements were performed at the Institut Laue-Langevin, Grenoble, France.
Polarised neutron diffraction experiments were carried out using the D7 instrument.\cite{2009_D7} 
D7 is a cold neutron diffuse scattering spectrometer equipped with $xyz$ polarisation analysis.
It has 3 banks of 44 $^3$He detectors, which cover an angular range of approximately 132$^{\circ}$.
The sample (1.6~g)  was fixed to an aluminium holder defining the horizontal ($h0l$) scattering plane.
We used a wavelength of 4.8~\AA, which gave a $Q$-range coverage of 0.09 to 2.5~\AA$^{-1}$.
Scans were made by rotating the sample around the vertical axis with a 1$^{\circ}$ step size.
Quartz was used to normalise the polarisation efficiency, while vanadium was used to normalise the detector efficiency of the instrument.  

Single crystal neutron diffraction in an applied magnetic field was carried out using the D10 diffractometer.
A two dimensional 80$\times$80~mm$^2$ area detector was used for all measurements.\cite{1988_Wilkinson}
An asymmetric cryomagnet with vertical opening of $+20^{\circ}$ and $-5^{\circ}$ supplying a magnetic field of up to 60~kOe was used throughout the experiment.
The sample (0.33~g) was fixed to an aluminium holder and the magnetic field was applied along the vertical $c$~axis defining the horizontal $(hk0)$ scattering plane.
The accuracy of the alignment of the $c$~axis of the crystal along the magnetic field was better than 0.7$^{\circ}$. 
Incident wavelength of $\lambda =2.36$~\AA\ and $\lambda =1.26$~\AA\ from a pyrolytic graphite (PG) and Cu(200) monochromator respectively were utilised.
The shorter wavelength was utilised in order to the reach the out of scattering plane incommensurate reflections.
A PG filter was used to reduce contamination of $\lambda/2$.
The field and temperature scans were performed by measuring the peak intensity while ramping the field/temperature.
The integrated peak intensity was found by summing the counts in a small area of the detector surrounding the reflection.
Magnetic refinements of the magnetic phases were carried out using the \textsc{fullprof} suite.\cite{Fullprof}

\begin{figure*}[t]
\begin{center}
\includegraphics[width=1.6\columnwidth]{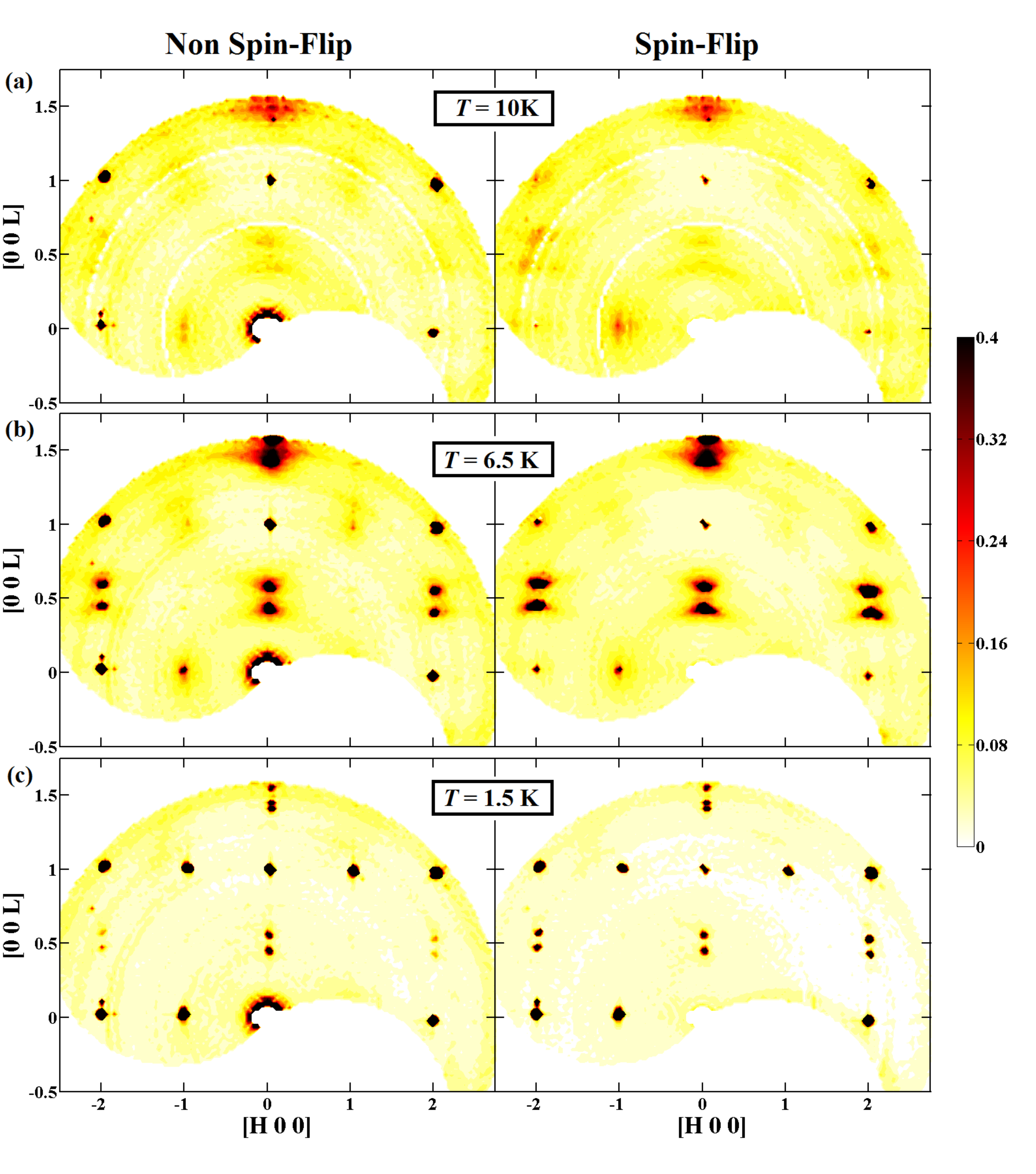}
\caption{(Colour online) Single crystal neutron diffraction maps of $(h0l)$ plane for \HoB\ measured using the D7 diffractometer.
The non spin-flip (left column) and spin-flip (right column) channels at different temperature are shown.
(a) 10~K, paramagnetic phase, (b) 6.5~K, IT incommensurate phase, (c) 1.5~K, low-temperature \afm\ phase.}
\label{fig: Z_pol_map}
\end{center}
\end{figure*}

A Quantum Design SQUID magnetometer was used to measure magnetic susceptibility, $\chi(T)$, and an Oxford Instruments vibrating sample magnetometer for the field dependent magnetisation.
Susceptibility measurements showed a broad maximum at \TNH\ and a small discontinuous drop at \TNL, consistent with previous measurements.\cite{2008_Okuyama_HoB4}
However, in fields below 1~kOe, there is an addition weak feature visible at $T_{\mathrm{C}}=15$~K for the crystal used in the D10 experiment.
This has been attributed to the ferromagnetic transition in a small HoB$_2$ impurity.\cite{2006_Roger_HoB2}
No feature in $\chi(T)$ nor depolarisation of the neutron beam was observed for the crystal used during the D7 experiment suggesting the ferromagnetic impurity was insignificant for this sample.

\section{Results and Discussion} \label{Sec_Results}
\subsection{Polarised Neutron Measurements} \label{SubSec_D7}

Fig.~\ref{fig: Z_pol_map} shows the zero-field intensity maps of the ($h0l$) scattering plane of \HoB.
For these maps a $z$-polarisation was used, which coincided with the vertical [010]~axis.
Two independent channels were measured, here called the non-spin flip (NSF) and the spin-flip (SF) channel.
The NSF channel is sensitive to nuclear scattering and scattering from a component of the magnetic moment parallel to the neutron polarisation, while the SF channel is only sensitive to the component of the magnetic moment in the scattering plane and perpendicular to the scattering vector, $Q$.
Hence, the sum of the NSF and SF channels gives the total intensity that would be observed in an unpolarised neutron experiment.
In addition there is a small degree of ``leakage'' between the two channels arising from imperfect polarisation of the neutron beam.
Although this is corrected for by normalising to a quartz sample, the correction does not work perfectly for high-intensity Bragg peaks giving rise to a small systematic error corresponding to the exact positions of nuclear Bragg peaks.
A comparison of the intensity shows these features are approximately 1\% of the intensity of the NSF counterparts and are easily distinguished from real features.

The paramagnetic phase at $T=10$~K, close to the first magnetic transition is shown in Fig.~\ref{fig: Z_pol_map}(a).
The nuclear Bragg reflections are observed in the NSF channel, while the weak ``leakage'' features are observed in the SF channel.
Broad magnetic features are observed around (0~0~0.43), (0~0~1.43), $(\bar{2}$~0~0.43) and (2~0~0.43) corresponding to the positions of incommensurate reflections appearing at lower temperatures.
The first transition occurs at \TNH\ and this is diffuse scattering arising from short-range correlations occurring above ordering. 
Additional diffuse magnetic features are observed at $(\bar{1} 0 0)$, $(1 0 0)$, $(\bar{1} 0 1)$, and $(1 0 1)$.
Again these arise due to short-range correlations forming before the onset of long range \afm\ order.

Fig.~\ref{fig: Z_pol_map}(b) shows the IT phase, with an incommensurate propagation vector of ($\delta, \delta, \delta^\prime$), where $\delta=0.02$ and $\delta^\prime=0.43$.
Although we are unable to resolve $\pm\delta$ separately using D7, we found this to be the propagation vector with the unpolarised neutron measurements on the D10 diffractometer (see below, Sec.~\ref{SubSec_field}).
We see a dramatic increase in the intensity of broad features at positions predicted by this propagation vector in both the NSF and SF channels.

This suggests that there is a component of the magnetic moment in the horizontal $(h0l)$ scattering plane as well as one parallel to the [010] direction  and is consistent with the proposed incommensurate magnetic structure~\cite{2008_Okuyama_HoB4}, which is described in more detail in the following section~\ref{SubSec_field}.
Additionally we see the appearance of weak features at $(\bar{1}00)$, $(101)$ and $(\bar{1}01)$ positions in the reciprocal space.
These reflections are associated with the commensurate \afm\ structure properly formed in the low temperature phase below \TNL, which is starting to appear at intermediate temperatures. 

The low temperature phase (Fig.~\ref{fig: Z_pol_map}(c)) shows a large increase in the intensity of the (100) reflection associated with the \afm\ order as well as an increase in the intensity of all the other commensurate reflections indicating a $\mathbf{q}=0$ \afm\ structure.
Surprisingly we also see the incommensurate reflections persist down to the lowest experimentally available temperature, 1.5~K, although with reduced intensity.

\begin{figure}[tb] 
\begin{center}
\includegraphics[width=0.95\columnwidth]{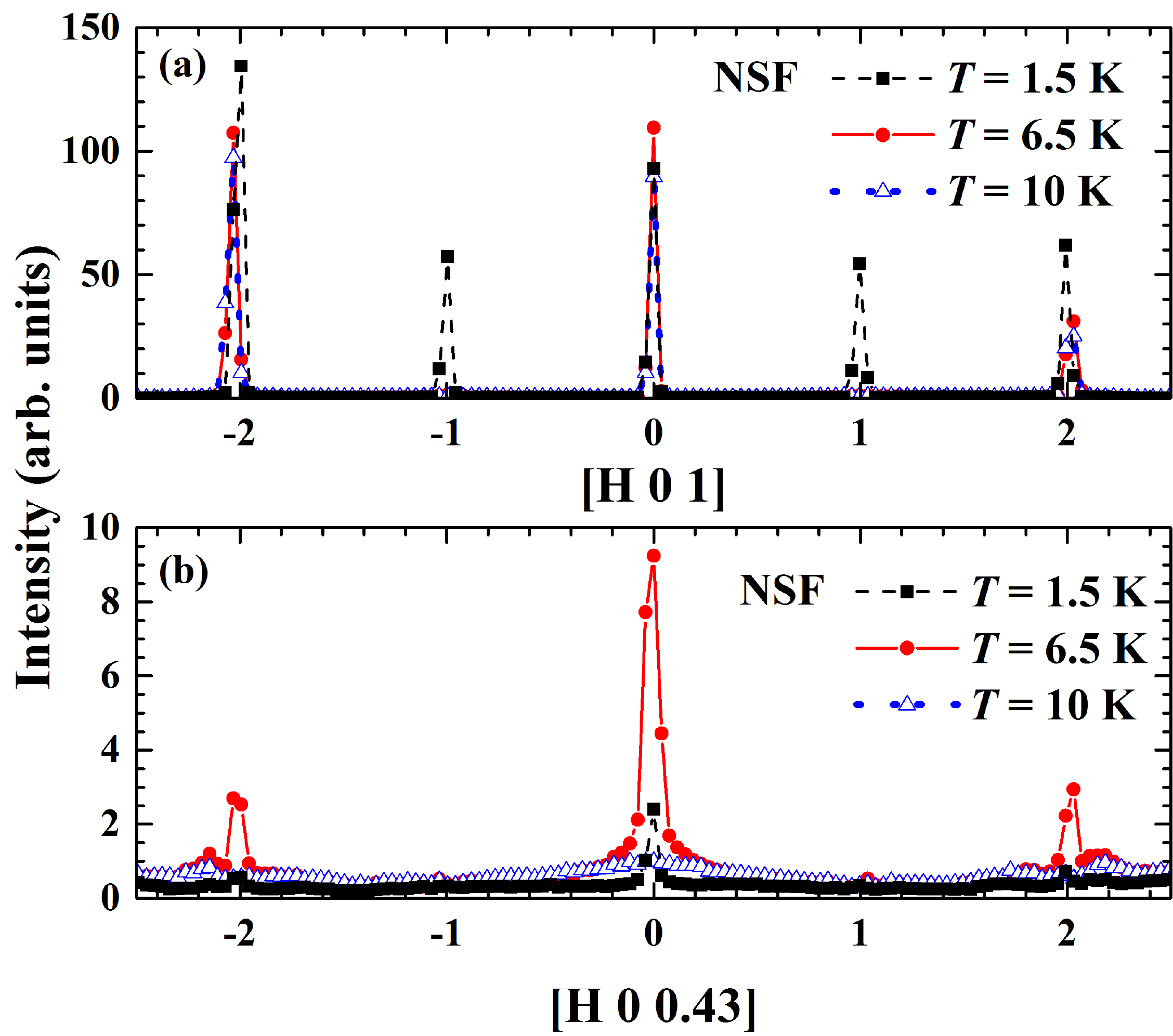}
\caption{(Colour online) Line of cuts of the $z$-polarisation intensity maps at 10, 6.5 and 1.5~K for the [H~0~L] direction, where (a) $\rm L=1.0 \pm 0.1$ and (b)  $\rm L=0.43 \pm 0.10$.}
\label{fig: line_cuts}
\end{center}
\end{figure}

We have taken line cuts through the reciprocal space maps along the [H~0~L] direction for the NSF channel, where $\rm L=0.43 \pm 0.10$ and $1.0 \pm 0.1$ to illustrate the temperature evolution of the scattering patterns.

Fig.~\ref{fig: line_cuts}(a) showing the cuts along [H~0~1] reveals an increase in intensity of the Bragg peaks between the paramagnetic and IT phases, with a further increase on the $(201)$ and $(\bar{2}01)$ reflections in the low temperature phase.
However, there is a decrease in the intensity of the (001) peak between 6.5 and 1.5~K suggesting that this peak is associated with the incommensurate magnetic structure and explains its presence in the SF channel at 1.5~K.  

Fig.~\ref{fig: line_cuts}(b) showing the line cut [H~0~0.43] shows three peaks at the incommensurate positions.
A broad diffuse scattering feature centred on (0~0~0.43) is clearly seen at 10~K.
At 6.5~K we see a well defined peak form on top of the diffuse scattering.
This two component aspect of the peak suggests there is some disorder associated with the incommensurate phase, which is consistent with previous measurements.\cite{2008_Okuyama_HoB4}
However we observe at 1.5~K the peaks persisting, with reduced intensity and the diffuse scattering has disappeared.
This can also be seen in the intensity maps (Fig.~\ref{fig: Z_pol_map}) where the broad features are well defined on cooling from 6.5 to 1.5~K in both the NSF and SF channels.
We therefore suggest that the remnant intensity arises due to the incommensurate magnetic structure becoming ``frozen'' at low temperatures.  

\subsection{Applied Field Measurements} \label{SubSec_field}
\begin{figure}[tb]
\begin{center}
\includegraphics[width=0.9\columnwidth]{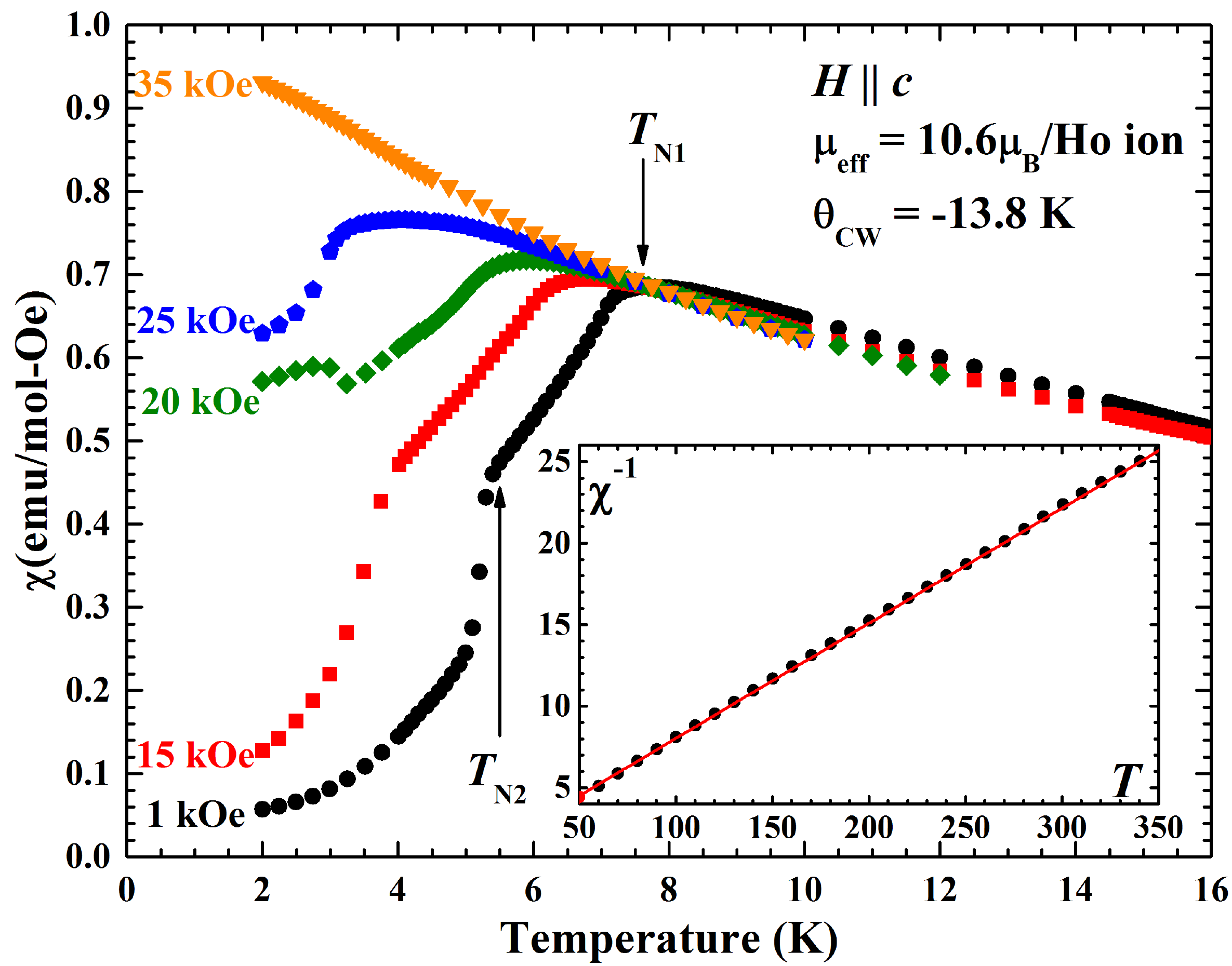}
\caption{(Colour online) Temperature dependent magnetic susceptibility in increasing fields applied along $H \parallel c$. Zero-field transition temperatures are indicated by the arrows.
				The inset shows the inverse susceptibility for $H=1$~kOe.}
\label{fig: chi_fields}
\end{center}
\end{figure}

Fig.~\ref{fig: chi_fields} shows the magnetic susceptibility in different fields.
The susceptibility curve at $H=1$~kOe exhibits a broad maximum at $T_{\mathrm{N1}}=7.1$~K and a discontinuous drop at $T_{\mathrm{N2}}=5.7$~K, reflecting the first-order nature of the phase transition between the IT and the low-temperature magnetic structures.~\cite{2008_Okuyama_HoB4}

The inverse susceptibility shows Curie-Weiss behaviour for $T>50$~K.
The effective magnetic moment was determined to be $\mu_{\mathrm{eff}}=10.6 \mu_{\mathrm{B}}$ per Ho ion and the Curie-Weiss constant $\theta_{\mathrm{CW}}=-13.8$~K.
These are all in agreement with previously published results.\cite{2009_Kim_HoB4}
Increasing the field suppresses the ordering temperature and $T_{\mathrm{N2}}$ is no longer present above 20~kOe, while $T_{\mathrm{N1}}$ persists up to at least 25~kOe.

\begin{figure}[tb]
\begin{center}
\includegraphics[width=0.9\columnwidth]{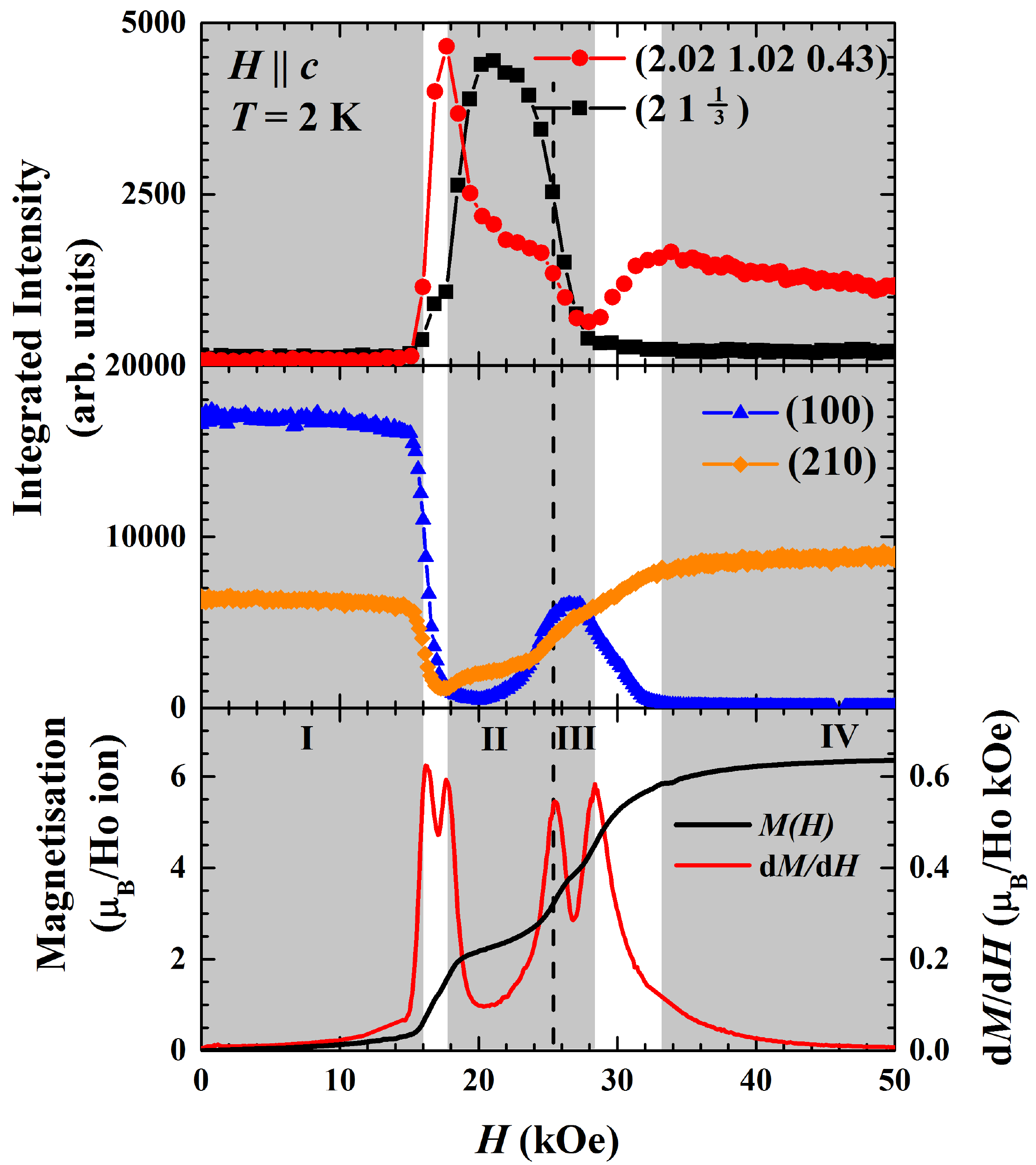}
\caption{Evolution of (top panel) fractional and (middle panel) integer $(hkl)$ reflections intensities with magnetic field compared to (bottom panel) the magnetisation curve for the magnetic field ramping up at $T=2$~K.
	There are four \emph{stationary} magnetic structures, coloured in grey and labelled as Phases~I, II, III and IV, in which magnetisation remains almost constant, while the white regions are \emph{transitionary} structures in which magnetisation is rapidly changing. The dashed line corresponds to a narrow transitionary state between phases II and III.}
\label{fig: 3_panel}
\end{center}
\end{figure}

Fig.~\ref{fig: 3_panel} compares the intensity of fractional $(2.02\,1.02\,0.43)$, $(h\,k\,\frac{1}{3})$ and integer $(hkl)$ reflections with the field dependent magnetisation curve at $T=2$~K.
Field dependent neutron and magnetisation measurements presented were performed by ramping the magnetic field up and down, however for clarity unless stated otherwise only results where the field was ramped up are shown.
The magnetisation curve shows a wide plateau at \MMF~$=\frac{1}{3}$, with two smaller features (best seen as local minima in the derivative of the magnetisation) occurring at \MMF~$\approx \frac{1}{6}$ and $\frac{3}{5}$.
The $\frac{1}{3}$ and $\frac{3}{5}$ have been observed in previously published results~\cite{2010_Matas, 2009_Kim_HoB4} whilst the feature at \MMF~$\approx\frac{1}{6}$ has not been observed previously.
The magnetisation saturates at approximately 6.5\muB\ per Ho ion, which is significantly smaller than 10\muB, the predicted saturation moment for a Ho$^{3+}$ ion.
It is presently unclear at what strength of an applied field the magnetisation will reach the value corresponding to the full polarisation of the magnetic moment.

Throughout this section the magnetic states will be referred to as either stationary states, where the magnetisation remains almost constant with increasing field and transitionary states, where there is a rapid change in the magnetisation.
The stationary magnetic phases observed in \HoB\ are labelled in Fig.~\ref{fig: 3_panel} by roman numerals from I to IV.
Phase~I corresponds to the long range non-collinear \afm\ order, found in the zero field.
The field dependence of the \afm\ (100) reflection shows a sharp decrease in intensity at 16~kOe coupled with a sharp increase in the intensity of the incommensurate $(2.02\,1.02\,0.43)$ reflection, as the \afm\ order gives way to the incommensurate magnetic state in a transitionary state between Phases~I and II.
In this transitionary state, there is also a gradual intensity increase in $(h\,k\,\frac{1}{3})$-type peaks which reach a maximum in Phase~II, coinciding with the $\frac{1}{3}$-magnetisation plateau.
Interestingly the incommensurate reflections persist in Phase~II as well, although with a reduced intensity.
Phase~III sees an increase in the (100) reflection, suppressing the incommensurate reflections, suggesting the onset of an \afm ally ordered phase.
The region above Phase~III is a transitory state, where there is a decrease in the intensity of (100) reflection and an increase in both the ferromagnetic (210) and incommensurate reflections up to 33~kOe.
The intensity of the ferromagnetic (210) peak plateaus in Phase~IV and the intensity of the incommensurate reflection gradually decreases with further field increase.
The shading used in  Fig.~\ref{fig: 3_panel} emphasises the difference between stationary (grey) and transitionary (white) states, while an additional narrow transitionary phase between II and III is indicated by a dashed line.

\begin{figure}[tb]
\begin{center}
\includegraphics[width=0.9\columnwidth]{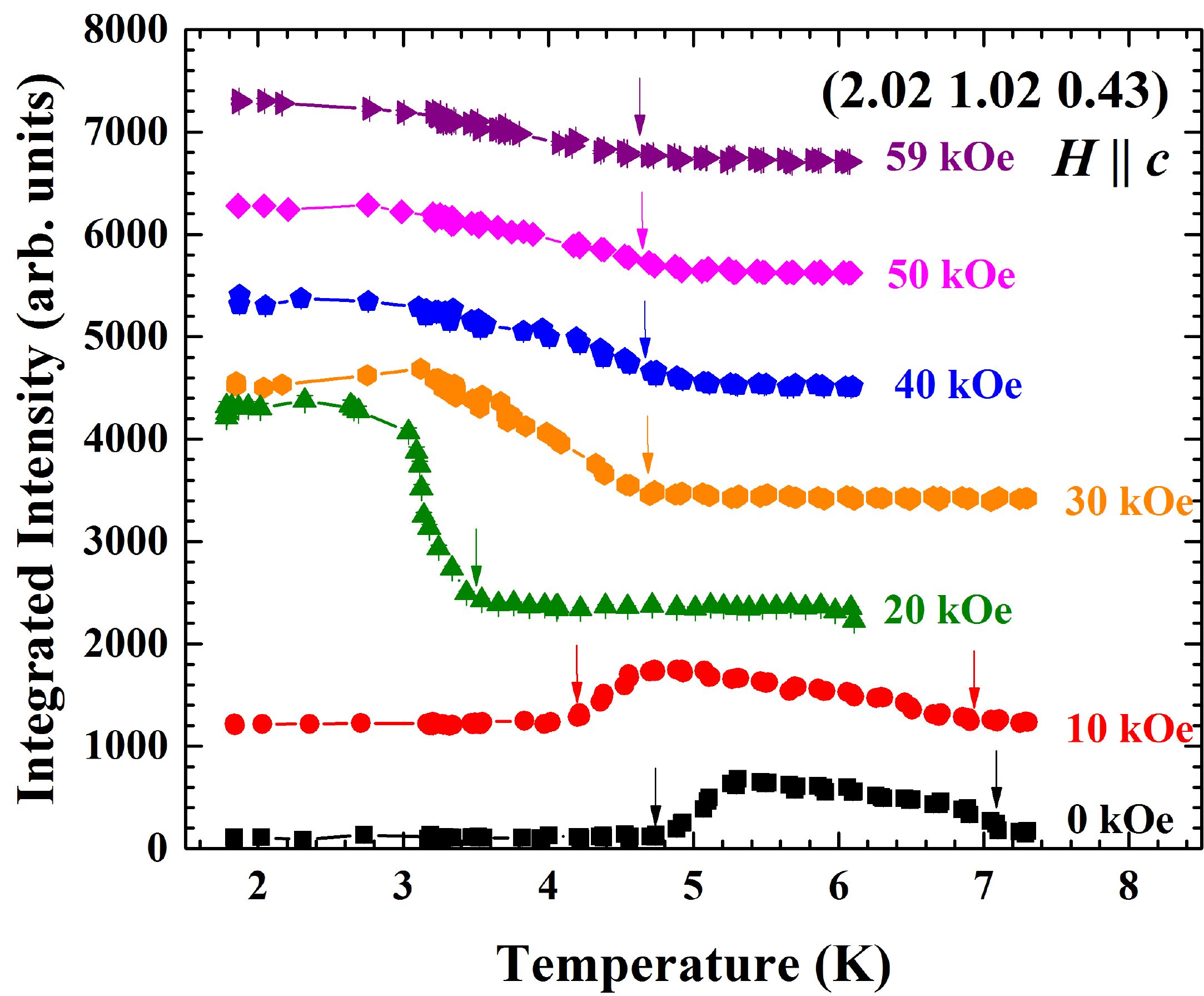}
\caption{Temperature dependence of the intensity incommensurate reflection $(2.02\,1.02\,0.43)$ in different fields.
	A stabilisation of the incommensurate phase is observed at 20~kOe, while a decrease of the low temperature intensity with increasing field.
	Each curve is sequentially offset by 1100 counts.
	}
\label{fig: 0p44_temp}
\end{center}
\end{figure}

The temperature dependence of the intensity of the $(2.02\,1.02\,0.43)$ reflection in different fields is shown in Fig.~\ref{fig: 0p44_temp}.
The zero field data shows the appearance of the $(2.02\,1.02\,0.43)$ reflection between $T_{\mathrm{N2}}<T<T_{\mathrm{N1}}$ corresponding to the incommensurate IT phase.
On increasing the field to 10~kOe, the incommensurate phase region shifts to lower temperatures.
Increasing further to 20~kOe, which is on the tail-end of the re-established phase we see a sharp transition with the appearance of the $(2.02\,1.02\,0.43)$ reflection at 3.5~K.
Increasing the field further up to 59~kOe we see the transition becomes much more gradual, starting at 4.8~K, with the intensity levelling off at approximately 3.2~K.
The levelled off intensity at low temperature then decreases with increasing magnetic field.
Whether phase~IV is an incommensurate phase or a polarised state with a ``frozen-in'' incommensurate component which is shrinking with increasing field cannot be determined from the present data and further investigation is required.

\begin{figure}[b]
\begin{center}
\includegraphics[width=0.95\columnwidth]{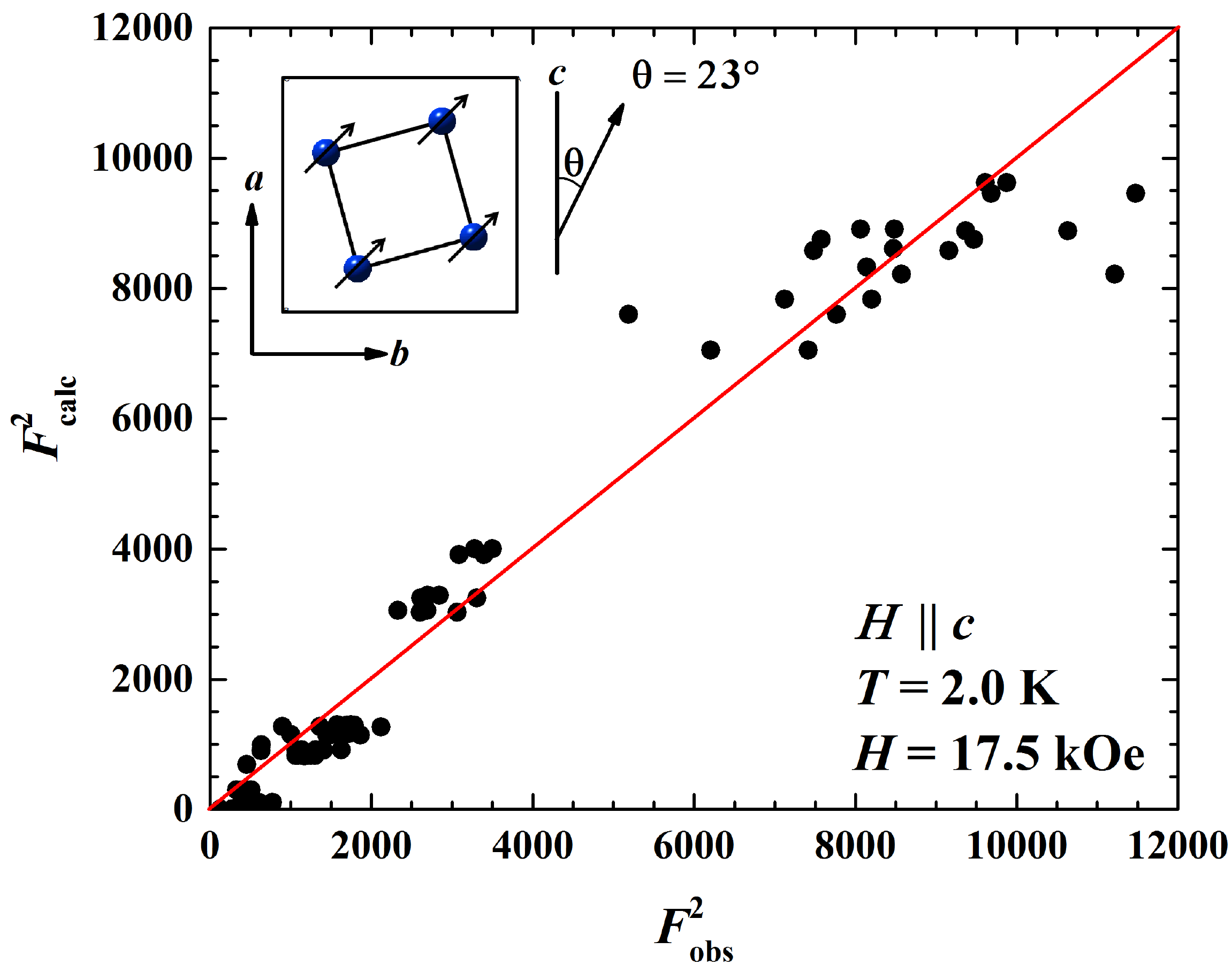}
\caption{Comparison of the calculated and observed intensity for the incommensurate phase re-established at $T=2.0$~K in an applied field of 17.5~kOe. The model for the zero-field incommensurate structure was used to calculate the intensity. }
\label{fig: 0p43_comp}
\end{center}
\end{figure}

In the transitionary state between Phases~I and II at $H = 17.5$~kOe we have collected the integrated intensities of a set of 120 incommensurate reflections and performed a magnetic refinements using the model for the zero field incommensurate structure.~\cite{2008_Okuyama_HoB4}
The fit ($R_{\mathrm{Bragg}}=15.84$\%) is shown in Fig.~\ref{fig: 0p43_comp}.
We confirmed there was a component of the magnetic moment in the $ab$~plane as well as along the $c$~axis.
The moments are tilted from the $c$~axis by approximately $23^{\circ}$  compared to $25^{\circ}$  degrees reported by Okuyama {\it et al.}~\cite{2008_Okuyama_HoB4} for zero field and the $ab$~plane component points along the [110] direction (see inset Fig.~\ref{fig: 0p43_comp}). 
There is an amplitude modulation, which is most prominent along the $c$~axis as a consequence of the propagation vector ($\delta$,~$\delta, $~$\delta'$) being close to commensurate position in the $ab$~plane.

\begin{figure}[tb]
\begin{center}
\includegraphics[width=0.95\columnwidth]{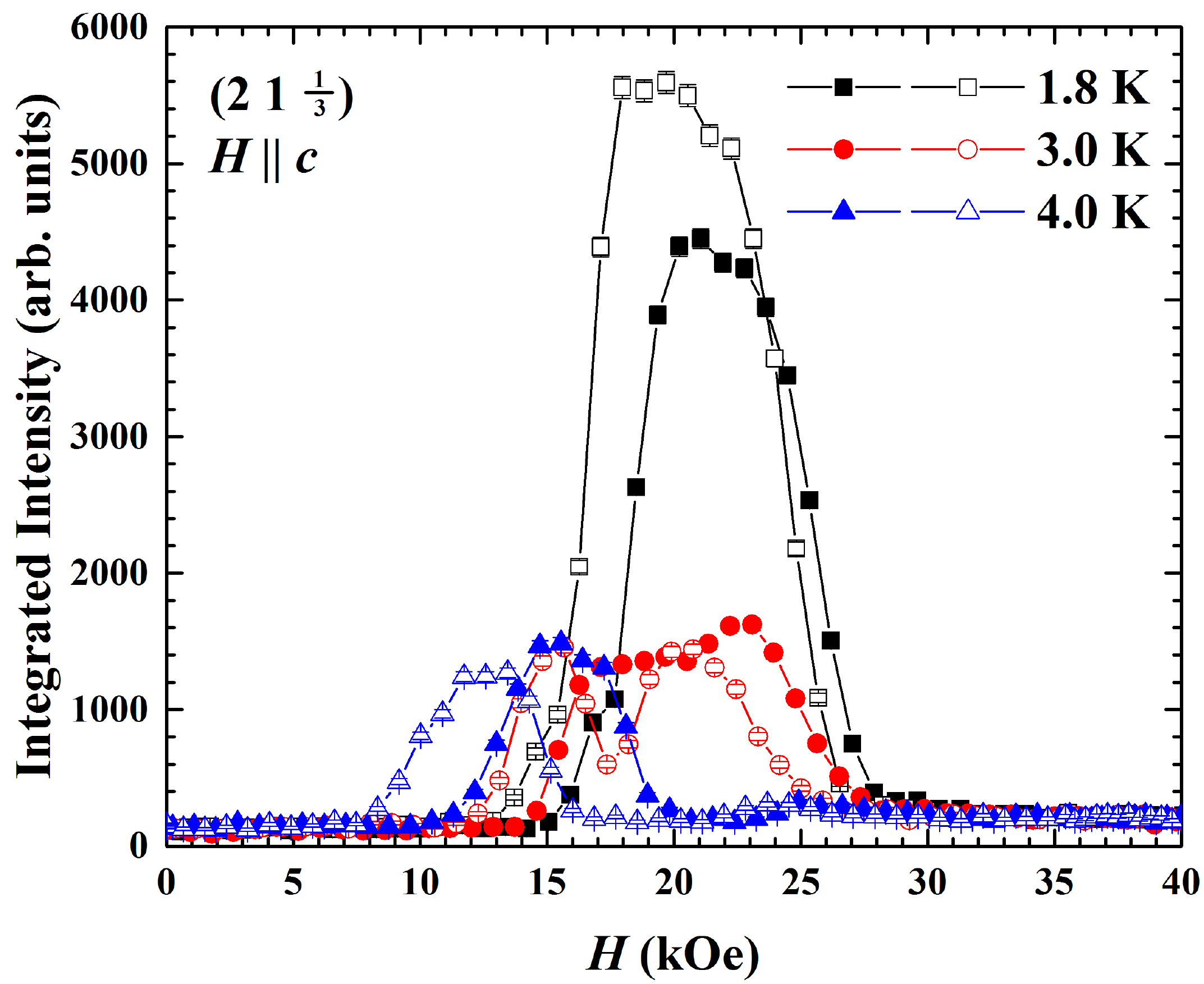}
\caption{Field dependence of the intensity of $(2\,1\,\frac{1}{3})$ reflection at different temperatures.
	Filled symbols correspond ramping the field up, empty symbols correspond to ramping the field down.
	The temperature increase between $T=1.8$ to 4~K causes a visible shift of the magnetic phase corresponding to $\frac{1}{3}$-magnetisation plateau to lower fields.}
\label{fig: 0p33_field}
\end{center}
\end{figure} 

\begin{figure}[tb]
\begin{center}
\includegraphics[width=0.97\columnwidth]{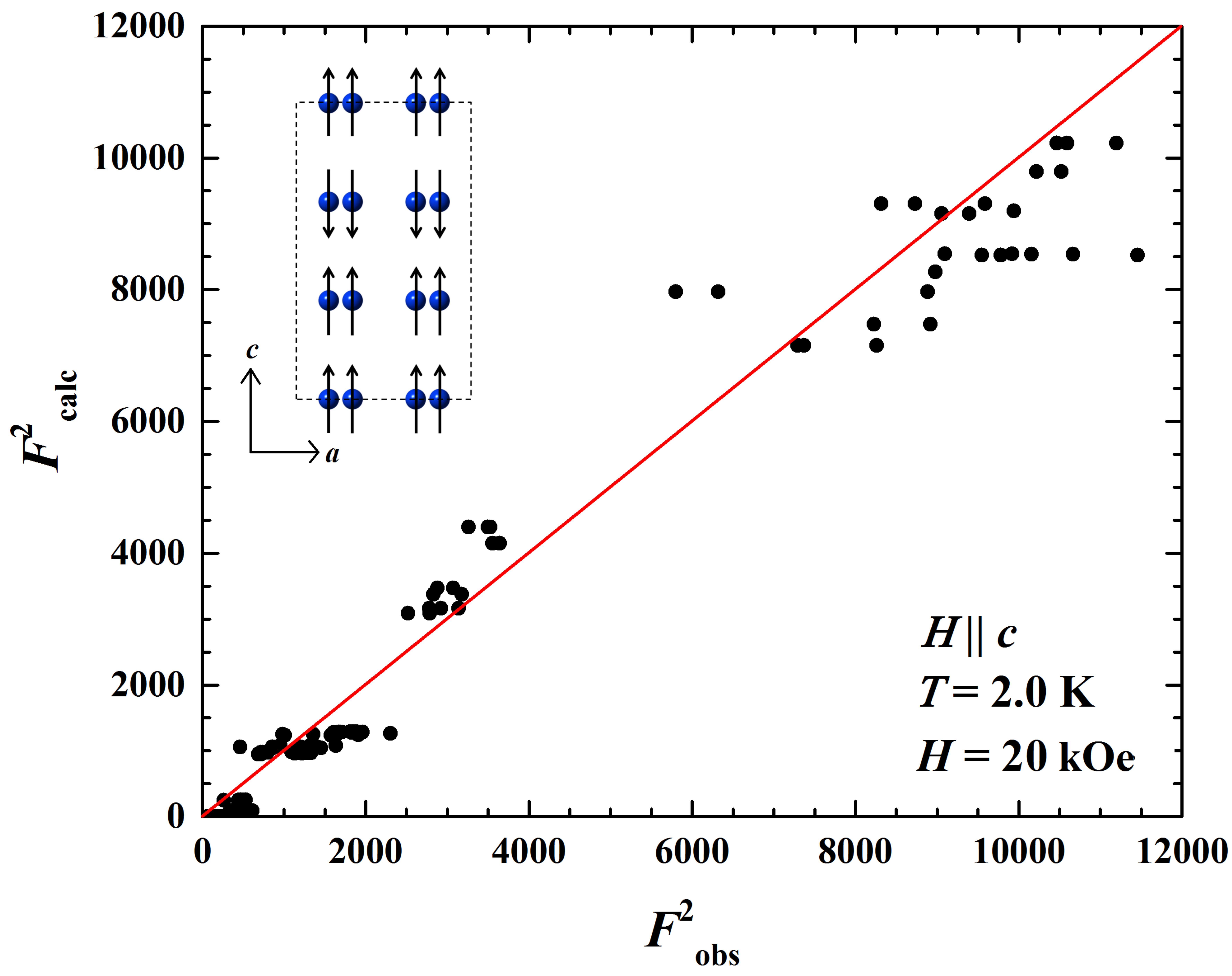}
\caption{Comparison of the calculated and observed intensity for Phase~II corresponding to the $\frac{1}{3}$-magnetisation plateau.
		An \emph{up-up-down} ferrimagnetic phase along the $c$~axis was used to calculate the intensity.
		The proposed magnetic structure is shown in the inset.	
}
\label{fig: 0p33_comp}
\end{center}
\end{figure}

Fig.~\ref{fig: 0p33_field} shows the field dependence of the intensity of the $(2\,1\,\frac{1}{3})$ reflection at different temperatures.
The nonzero intensity of this and other symmetry related $(h\,k\,\frac{1}{3})$-type Bragg peaks marks the presence of the Phase~II coinciding with the $\frac{1}{3}$-magnetisation plateau. 
With temperature increasing from 1.8 to 4.0~K, the intensity of the $(2\,1\,\frac{1}{3})$ peak falls dramatically.
An increase in the temperature also results in a shift of the Phase~II to a lower field range.
In order to determine the magnetic structure of Phase~II we derived the irreducible representations for a propagation vector of (0, 0, $\frac{1}{3}$). 
Using a collection of 127 $(h\,k\,\frac{1}{3})$-type reflections, we found the best fit for Phase~II arises from a basis vector with the moments aligned parallel to $c$~axis.
The moments form ferromagnetic layers in the $ab$-plane, which stack in an \emph{up-up-down} arrangement, expanding the crystallographic unit cell along the c axis by a factor of three.
All the moments have equal magnitude with one of the three planes in the unit cell pointing anti-parallel compared to the other two layers, thus there is a net magnetisation along the c axis which is $\frac{1}{3}$ compared to the full ferromagnetic magnetisation.
Many of the proposed structures for the $\frac{1}{3}$ and other fractional magnetisation plateaux in the \RB\ family involve the arrangement of the magnetic moments only within a single plane of $R$ ions, while the $\frac{1}{3}$-magnetisation plateau in \HoB\ is clearly caused by the ferrimagnetic stacking of the planes.
While there are a number of additional interactions to nearest neighbour and next nearest neighbour are included in these theoretical models (e.g. RKKY interactions, quadrupolar, etc.), with the present results we are unable to clarify the interactions involved in the formation of the UUD structure.
The comparison of the observed and calculated intensity for this fit ($R_{\mathrm{Bragg}}=15.21$\%) is shown in Fig.~\ref{fig: 0p33_comp}.
The refinements gave a $c$-component of the magnetic moment to 1.95\muB\ per Ho ion, which is in agreement with the value observed in field dependent magnetisation data (Fig.~\ref{fig: 3_panel}).
However, this observation does not rule out the presence of significant magnetisation component orthogonal to the $c$ axis both in Phase~II and Phase~IV.

As can be seen in Fig.~\ref{fig: 3_panel}, the $(100)$ reflection present at low temperature in lower fields (Phase~I) initially disappears in an applied field of about 16~kOe and then it is re-established in higher fields (Phase~III) with a maximum intensity seen around 27~kOe.
This reflection is not allowed by the tetragonal crystal symmetry and its presence indicates an \afm\ ordering.
We have attempted a magnetic structure determination of this \afm\ phase and for these purposes have subtracted the high temperature (30 K), zero-field intensity of a full collection of integer $(hkl)$ reflections from those at 27~kOe to obtain the purely magnetic intensity.
On inspection, the intensity of some reflections, such as the $(140)$ and $(330)$ and their Friedel pairs, has actually decreased on cooling down to base temperature and increasing the field to 27~kOe, therefore the subtraction returned a negative intensity for them.
Due to a significant net magnetisation in this field we would expect the intensity of the $(140)$ and $(330)$ to either remain constant or for there to be an increase. 
The observed decrease in intensity could be due to a structural phase transition.
In order to test this hypothesis we applied a small distortion to the atoms in the unit cell, assuming the same symmetries.
For more information please refer to the supplementary material~\cite{Note1}.
We found a displacement of the Ho ions or a compression/expansion of the B octahedra predicted a decrease in intensity for both the (330) and (140) reflections.
Although we are unable to confirm a structural distortion with the current data, it is a possible explanation to the negative intensity observed after the subtraction.

\begin{figure}[tb]
\begin{center}
\includegraphics[width=0.95\columnwidth]{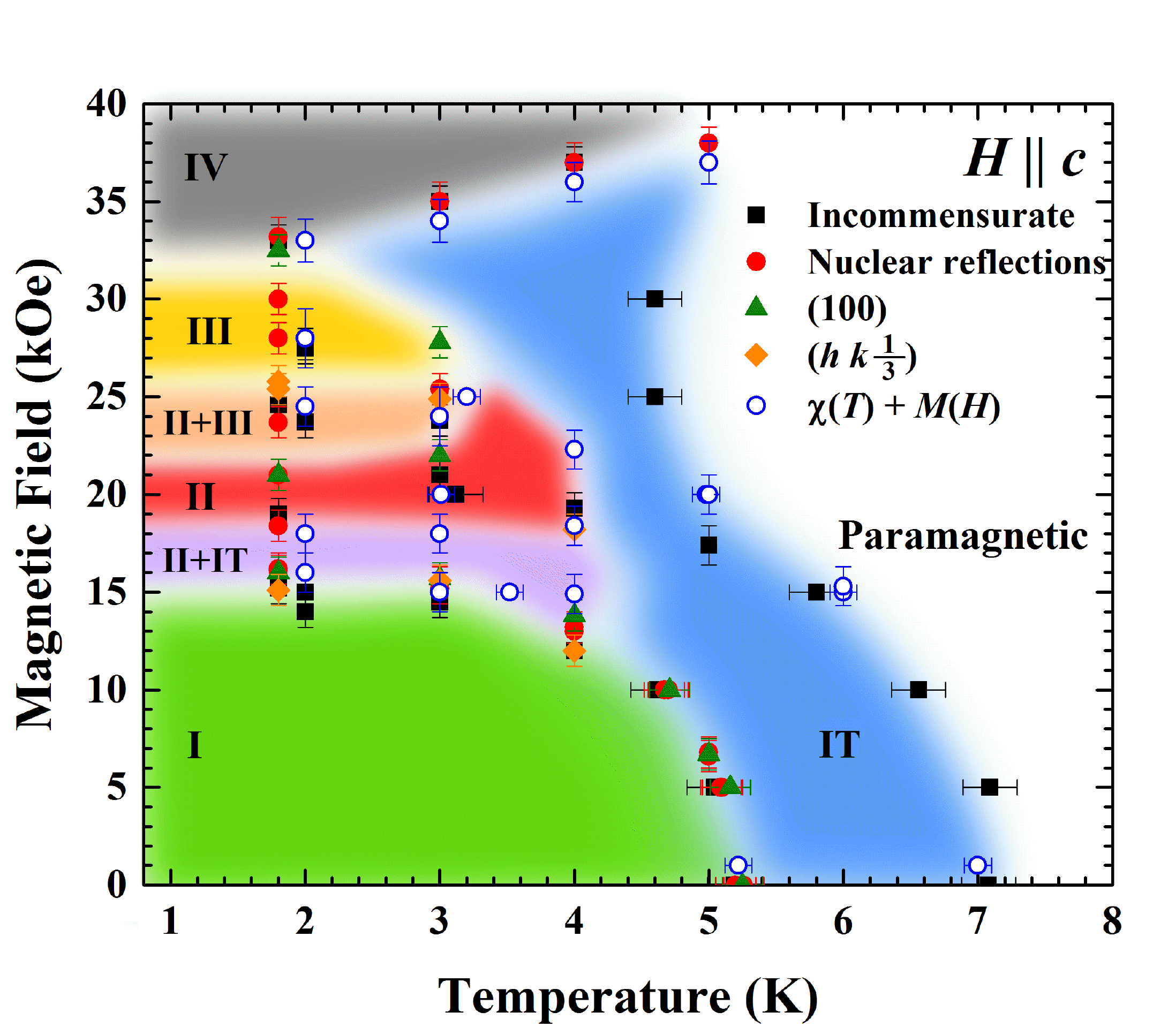}
\caption{(Colour online) Magnetic phase diagram of \HoB\ constructed from neutron diffraction data for different reflections (squares, filled circles, triangles and diamonds) and magnetisation (empty circles) measurements. All field dependent measurements were made by ramping the field up.
The labelling of different magnetic structures is consistent with the one used in Fig.~\ref{fig: 3_panel} for $T=2$~K, however, with the increased temperature the magnetic phases tend to overlap.
The separation between the stationary and transitionary phases also becomes less obvious on heating, with the \emph{transitionary} Phases~II+IT and II+III occupying large portions of the phase diagram.}
\label{fig: phase_dia}
\end{center}
\end{figure}

Finally we have constructed the magnetic phase diagram for $H \parallel c$ using temperature and field dependence of the intensity of different reflections from the single crystal neutron experiment as well as the magnetic susceptibility and magnetisation measurements.
The phase diagram is shown in Fig.~\ref{fig: phase_dia}.
The full set of experimental data used to construct the phase diagram is given in the supplementary material~\cite{Note1}.
Besides the paramagnetic regime and the IT structure, there are four stationary magnetic phases at low temperatures separated by mixed transitionary states.  
In zero-field there are two transitions $T_{\mathrm{N1}}$ and $T_{\mathrm{N2}}$ leading to an incommensurate magnetic IT state and a non-collinear \afm\ state (Phase~I) respectively. 
Application of a magnetic field suppresses both $T_{\mathrm{N1}}$ and $T_{\mathrm{N2}}$, while for a narrow field range ($16 < H < 18$~kOe) at low temperatures, the IT incommensurate structure acts as an intermediary phase between the non-collinear \afm\ Phase~I and the ferrimagnetic \emph{up-up-down} structure formed in Phase~II.
The magnetic structure of Phase~III remains undetermined, however the presence of the (100) reflection in the single crystal neutron data indicated that it corresponds to an \afm\ arrangement of the Ho moments in the $ab$~plane.
The state above Phase~III is an intermediary phase, where there is an increase in the intensity of the reflections corresponding to a ferromagnetic phase, suggesting that the moments are tilting towards the magnetic field direction.
While Phase~IV corresponds to a fully polarised phase with a ``frozen-in'' incommensurate state.     

\section{Summary} \label{Sec_Conclusions}
We have probed the field induced magnetic states of \HoB\ using single crystal neutron diffraction as well as magnetisation measurements.
Polarised neutron experiments in zero field have revealed diffuse scattering in the paramagnetic regime, heralding the onset of magnetic order.
Line cuts through the incommensurate reflection at (0~0~0.43) at $T$~=~6.5~K have shown the peak consists of two components suggesting some disorder in the incommensurate phase. 
These incommensurate reflections persist down to 1.5~K where a non-collinear \afm\ state is formed. 
For a magnetic field applied along the $c$~axis, four stationary phases were found, separated by intermediary states which are established over a narrow field ranges.
The non-collinear antiferromagnet established in zero field is present up to 16~kOe before the incommensurate state is re-established as the dominant phase between 16 and 18~kOe.
The magnetisation curve then shows a plateau at \MMF~=~$\frac{1}{3}$, coupled with the appearance of $(h\,k\,\frac{1}{3})$-type reflection arising from an \emph{up-up-down} ferrimagnetic structure along the $c$~axis.
An \afm\ phase is established between 21 and 33~kOe.
Finally there is a polarised phase with remnants of the incommensurate reflections which persist up to 59~kOe.
Whether this is an incommensurate phase or a polarised state with the zero field incommensurate phase frozen, can not be determined and further high field magnetisation measurements would be an interesting extension to this work.
We hope this study of the field induced magnetic states of \HoB\ will contribute towards the development of theoretical models to describe the plateaux observed in the $R$B$_4$ family. 

\begin{acknowledgments}
The authors acknowledge enlightening discussions with Pinaki Sengupta, Stephen Lovesey, Dmitry Khalyavin, Denis Golosov, and Frederic Mila, as well as financial support from the EPSRC, UK, through Grant EP/M028771/1.
We also thank T.E.~Orton for invaluable technical support.  
\end{acknowledgments}

\bibliography{References2}
\end{document}